\documentclass[10pt,conference]{IEEEtran}

\usepackage{graphicx}% Include figure files
\usepackage{dcolumn}% Align table columns on decimal point
\usepackage{bm}% bold math
\usepackage{xcolor}
\usepackage{qcircuit}
\usepackage{braket}
\usepackage{float}
\usepackage{algorithm}
\usepackage{algorithmicx}
\usepackage{algpseudocode}
\usepackage{mathtools}
\usepackage{dirtytalk}
\usepackage{hyperref}
\usepackage{amsmath}
\usepackage{amsfonts}
\usepackage{amssymb}
\usepackage{amsthm}
\usepackage{newlfont}
\usepackage{array}
\usepackage{cite}

\usepackage{physics}
\usepackage{siunitx}
\usepackage[T1]{fontenc}

\usepackage{setspace}

\algnewcommand{\algorithmicand}{\textbf{ and }}
\algnewcommand{\algorithmicor}{\textbf{ or }}
\algnewcommand{\OR}{\algorithmicor}
\algnewcommand{\AND}{\algorithmicand}

\providecommand{\keywords}[1]
{
  \textbf{\textit{Keywords---}} #1
}

\title{A Hybrid Quantum-classical Fusion Neural Network to Improve Protein-ligand Binding Affinity Predictions for Drug Discovery}
\author{%
L. Domingo\textsuperscript{1$\star$}, M. Chehimi\textsuperscript{1}, S. Banerjee\textsuperscript{2$\dag$}, S. He Yuxun\textsuperscript{2$\dag$}, S. Konakanchi\textsuperscript{2$\dag$}, L. Ogunfowora\textsuperscript{2$\dag$}, S. Roy\textsuperscript{2$\dag$},\\ S. Selvarajan\textsuperscript{2$\dag$}, M. Djukic\textsuperscript{1}, C. Johnson\textsuperscript{1}
\\
\IEEEauthorblockA{\textsuperscript{1}Ingenii Inc., New York, USA}
\IEEEauthorblockA{\textsuperscript{2}Purdue University, The Data Mine, USA}
\IEEEauthorblockA{\textsuperscript{\dag}All Purdue University authors have equally contributed}
\IEEEauthorblockA{$^\star$Corresponding Author, email: laia@ingenii.dev}\vspace{-1cm}
}\vspace{-1cm}

\begin{document}
\maketitle
\begin{abstract}
Drug discovery hinges on the accurate prediction of binding affinity between prospective drug molecules and target proteins that influence disease progression, which is financially and computationally demanding. Although classical and hybrid quantum machine learning models have been employed in previous studies to aid in binding affinity prediction, they encounter several issues related to convergence stability and prediction accuracy. In this regard, this paper introduces a novel hybrid quantum-classical deep learning model tailored for binding affinity prediction in drug discovery. Specifically, the proposed model synergistically integrates 3D and spatial graph convolutional neural networks within an optimized quantum circuit architecture. Simulation results demonstrate a 6\% improvement in prediction accuracy relative to existing classical models, as well as a significantly more stable convergence performance compared to previous classical approaches. Moreover, to scalably deploy the proposed framework over today's \emph{noisy intermediate-scale quantum (NISQ)} devices, a novel quantum error mitigation algorithm is proposed. This algorithm outperforms existing techniques and is capable of mitigating errors with gate noise probabilities, $p\leq 0.05$, while resulting in no additional overhead during the training and testing phases.
\end{abstract}


\IEEEpeerreviewmaketitle
\vspace{-0.1cm}
\section{Introduction}\vspace{-0.05cm}
\label{sec:intro}
\indent Drug discovery relies on precise computational modeling of compound molecular interactions and understanding the role of proteins in disease mechanisms. Identifying proteins that drive molecular interactions leading to specific diseases is crucial. Once a target protein is identified, prospective drug candidates, often small molecules or \emph{ligands}, are generated. Ideal ligands are chosen based on high binding affinity to the target protein and minimal off-target interactions \cite{quantummechanical}. However, quantifying such binding affinities is computationally and financially demanding \cite{ddPNAS}.
%Drug discovery relies on the precise computational modeling of compound molecular interactions and understanding the role of proteins in disease mechanisms. Particularly, it is imperative to identify proteins that are instrumental in the cascade of molecular interactions leading to a specific disease. Upon the identification of such a target protein, a list of prospective drug candidates is generated. These candidates, often described as small molecules or compounds termed \emph{ligands}, have the potential to modulate the target protein's activity through binding interactions~\cite{quantummechanical}. Ideal ligands are chosen based on their high binding affinity to the target protein, coupled with minimal off-target interactions with other proteins. However, quantifying such binding affinities is a computationally and financially demanding \cite{ddPNAS}.
The transition from laboratory methods to computer-aided design (CAD) has markedly improved drug discovery efficiency. This advancement has been bolstered by artificial intelligence (AI) and machine learning (ML) algorithms, which facilitate exhaustive analyses of large-scale datasets, uncovering patterns related to atomic features of protein-ligand molecular complexes \cite{vamathevan2019applications}. Furthermore, recent strides in quantum computing have added another layer of sophistication by offering unprecedented parallelized computational capabilities \cite{biamonte2017quantum}. Quantum machine learning (QML) models, in particular, manage the challenges of exponentially increasing data dimensionality, often outperforming traditional ML models under specific conditions \cite{debenedictis2018future,chehimi2024federated}. These advancements make QML and hybrid quantum-classical models highly promising for drug discovery~\cite{batra2021quantum}. The main challenge for those QML models is the noise prevalent in today's \emph{noisy intermediate-scale quantum (NISQ)} computers, which requires advanced quantum error mitigation techniques to yield scalable and accurate performance.
%The transition from conventional laboratory methods to computer-aided design (CAD) has markedly improved the efficiency and accuracy of drug discovery and binding affinity prediction. This advancement has been further bolstered by the incorporation of artificial intelligence (AI) and machine learning (ML) algorithms, which facilitate exhaustive analyses of large-scale datasets, uncovering previously undetected patterns related to the atomic features of protein-ligand molecular complexes \cite{vamathevan2019applications}. Furthermore, recent strides in quantum computing have added another layer of sophistication to drug discovery efforts, offering unprecedented parallelized computational capabilities \cite{cao2018potential}. Quantum machine learning (QML) models, in particular, are well-suited to manage the challenges of exponentially increasing data dimensionality, often outperforming traditional ML models under specific conditions \cite{debenedictis2018future,chehimi2024federated,biamonte2017quantum,chehimi2023foundations}. Taken together, these technological advancements make QML and hybrid quantum-classical models highly promising for navigating the complex, high-dimensional challenges intrinsic to drug discovery~\cite{batra2021quantum}. The main challenge facing such QML models is the noise prevalent in today's \emph{noisy intermediate-scale quantum (NISQ)} computers, which requires advanced quantum error mitigation techniques to yield a scalable and accurate performance.

\begin{figure*}[t!]
    \centering
    \includegraphics[width=0.9\textwidth]{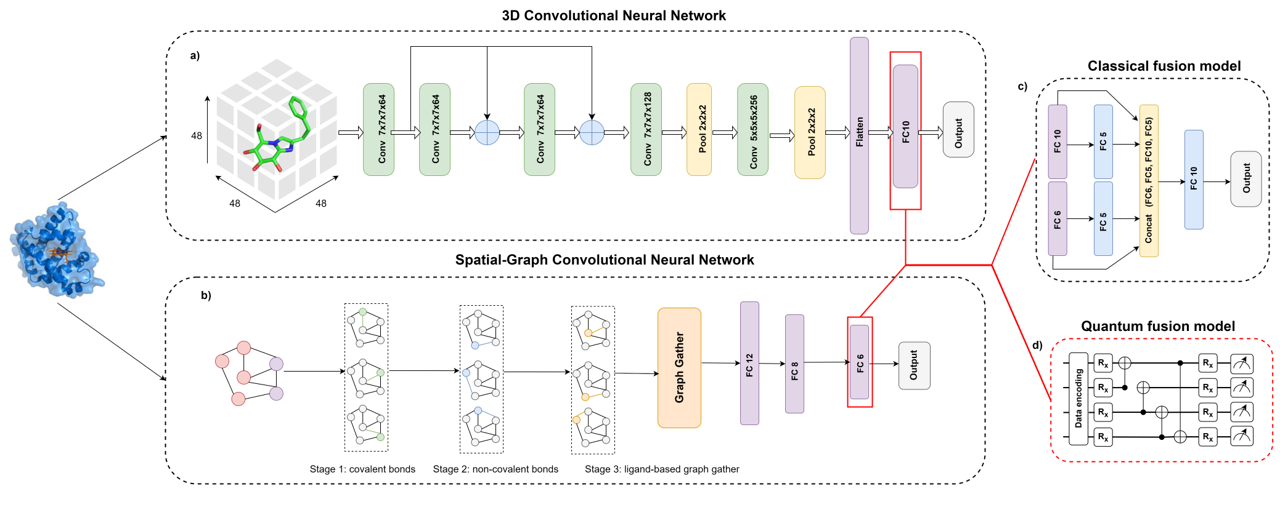}\vspace{-0.5cm}
    \caption{a) 3D-CNN architecture, b) SG-CNN architecture, c) Classical fusion model used as a benchmark, d) Proposed hybrid quantum fusion model. In the classical and hybrid fusion models, outputs from the second-to-last layer of each of the 2 classical neural networks trained with the PDBbind dataset are encoded onto quantum states before being fed into a feed-forward QNN.}\vspace{-0.5cm}
    \label{fig:Fusion model}
\end{figure*}

\paragraph*{Related Works}
Several prior works \cite{jimenez2018k,kwon2020ak,ATOM,mensa2023quantum,domingo2023hybrid} addressed binding affinity prediction in drug discovery using classical ML and QML. The work in \cite{jimenez2018k} leveraged 3D-convolutional neural networks (3D-CNNs) for protein-ligand binding affinity predictions. The work in \cite{kwon2020ak} enhanced \cite{jimenez2018k} by predicting binding affinities using an ensemble of independently-trained 3D-CNN network layers. The work in~\cite{ATOM} introduced a classical fusion model combining a 3D-CNN and a spatial graph CNN (SG-CNN), enhancing binding affinity prediction accuracy by concurrently processing grid-based, context-based, and graph-based protein features. However, these models lacked stable convergence and optimal predictive accuracy.
%Several prior works \cite{jimenez2018k,kwon2020ak,ATOM,mensa2023quantum,domingo2023hybrid} addressed the problem of binding affinity prediction in drug discovery using tools from both classical ML and QML. For instance, the work in \cite{jimenez2018k} leveraged 3D-convolutional neural networks (3D-CNNs) to perform protein-ligand binding affinity predictions in a faster and more efficient manner relative to other ML models. Moreover, the work in \cite{kwon2020ak} enhanced the model proposed in \cite{jimenez2018k} by predicting binding affinities using an ensemble of several independently-trained 3D-CNN network layers. Furthermore, the work in~\cite{ATOM} introduced a classical fusion model combining a 3D-CNN and a spatial graph CNN (SG-CNN). The model in~\cite{ATOM} enhances the binding affinity prediction accuracy by concurrently processing grid-based, context-based, and graph-based protein features. However, the models proposed in~\cite{jimenez2018k,kwon2020ak,ATOM} do not have a stable convergence, and their predictive accuracy is not optimized.
Additionally, the work in~\cite{mensa2023quantum} used quantum support vector machines for virtual drug screening, outperforming classical counterparts but focusing on a limited subset of protein-ligand features, restricting model accuracy and scalability. The work in~\cite{domingo2023hybrid} proposed a hybrid QML architecture, modifying a classical CNN by replacing a classical convolutional layer with an optimized quantum circuit, addressing computational complexity but not improving binding affinity predictive accuracy, achieving performance comparable to classical models. Moreover, these works did not analyze how the presence of noise in practical NISQ devices impacts the performance of the quantum solution. To the best of our knowledge, no existing research has effectively tackled the issue of binding affinity prediction in drug discovery by capitalizing on the benefits of QML while proposing error mitigation techniques to overcome practical noise in NISQ devices.
%Additionally, the work in~\cite{mensa2023quantum} uses quantum support vector machines for virtual drug screening. Although the hybrid QML model in~\cite{mensa2023quantum} outperforms classical counterparts, its focus on a limited subset of protein-ligand features restricts the model accuracy and scalability to larger datasets. Finally, the work in~\cite{domingo2023hybrid} proposes a hybrid QML architecture, modifying a classical CNN by replacing a classical convolutional layer with an optimized quantum circuit. While the work in~\cite{domingo2023hybrid} successfully tackles the computational complexity inherent in classical neural networks, it does not yield an improvement in binding affinity predictive accuracy, thus achieving a comparable performance to classical models. On top of that, prior works \cite{mensa2023quantum} and \cite{domingo2023hybrid} did not analyze how the presence of noise in practical NISQ devices impacts the performance of the quantum solution. To the best of our knowledge, no existing research has effectively tackled the issue of binding affinity prediction in drug discovery by capitalizing on the benefits of QML while simultaneously achieving high accuracy, ensuring smooth and stable convergence, and proposing error mitigation techniques to overcome practical noise in NISQ devices.

%our approach offers a targeted error mitigation strategy designed specifically for QNNs.
\paragraph*{Contributions}
The main contribution of this paper is the development of a novel hybrid quantum fusion model for enhancing binding affinity prediction in drug discovery. The model integrates a 3D-CNN and an SG-CNN, leveraging their strengths in processing diverse training data facets. The proposed quantum architecture is meticulously designed to ensure high accuracy. Simulation results demonstrate the superior performance of the proposed hybrid quantum fusion model relative to state-of-the-art classical models, achieving a 6\% improvement in binding affinity prediction accuracy and exhibiting faster, smoother, and more stable convergence. To enhance scalability on NISQ devices, we present a novel error mitigation technique capable of effectively alleviating noise introduced in quantum circuits for error probabilities $p\leq 0.05$ without additional quantum overhead during training and testing phases.
%The main contribution of this paper is the development of a novel hybrid quantum fusion model aimed at enhancing the binding affinity prediction in drug discovery. The proposed model strategically integrates a 3D-CNN and a SG-CNN to leverage their respective strengths in processing diverse facets of the training data. The proposed quantum architecture is meticulously designed to ensure high accuracy. Simulation results demonstrate the superior performance of the proposed hybrid quantum fusion model relative to state-of-the-art classical models. Particularly, the proposed model achieves a 6\% improvement in the binding affinity prediction accuracy, and exhibits faster, smoother, and more stable convergence, thereby boosting its generalization capacity. To enhance the scalability of the proposed quantum fusion model on NISQ devices, we present a novel error mitigation technique capable of effectively alleviating noise introduced in quantum circuits for error probabilities $p\leq 0.05$. This method incurs no additional overhead during both the training and testing phases, paving the way for broader QML applications over NISQ devices.

%The rest of this paper is organized as follows. Section \ref{Problem_setup} describes the proposed hybrid quantum fusion model, its components, the data used in training the model, and the necessary preprocessing steps. In Section \ref{experiments}, we conduct the experiments and simulations, and discuss key results and findings. Finally, conclusions are drawn in Section \ref{conclusion}.

\vspace{-0.3cm}
\section{System Model}\vspace{-0.2cm}
\label{Problem_setup}
This section describes the proposed hybrid quantum fusion model, its components, the data used in training the model, and the necessary pre-processing steps.

\vspace{-0.25cm}
\subsection{Proposed Hybrid QML Architecture}\vspace{-0.1cm}
 \label{classical_fusion}
The proposed hybrid quantum fusion architecture, shown in Fig.~\ref{fig:Fusion model}, integrates quantum neural networks (QNNs) into the the classical fusion model introduced in~\cite{ATOM}. In particular, the protein-ligand complex data (see Section \ref{Data}) is initially fed into a 3D-CNN and an SG-CNN, simultaneously. Then, late-late fusion is performed by feeding the outputs from the second-to-last layer of each of the two respective CNNs into a quantum fusion model, which incorporates a QNN. This proposed model is benchmarked against a classical fusion model, illustrated in Fig. \ref{fig:Fusion model}, which includes two feed-forward layers followed by a concatenation layer and two subsequent feed-forward layers, encompassing a total of 401 training parameters. Let us now briefly introduce the individual components of the proposed architecture in Fig.~\ref{fig:Fusion model}.

%\CR{fusion model. The classical fusion model, illustrated in Fig. \ref{fig:Fusion model}, includes two feed-forward layers with 5 neurons each, a concatenation layer, and two subsequent feed-forward layers with 10 and 1 neurons respectively, encompassing a total of 401 training parameters.}

\paragraph*{3D-CNN}
 \label{3D_CNN}
The adopted 3D-CNN model in this work is based on the ResNet architecture with two residual blocks~\cite{he2016deep}, mirroring the configuration detailed in \cite{ATOM}, facilitating comparison. The network contains five 3D convolutional layers, with 64, 64, 64, 128 and 256 filters respectively. All layers, except the final one, employ a kernel size of 7, with the last layer using a kernel size of 5. Two residual connections are incorporated to enable gradient passage to subsequent layers without a nonlinear activation function. Batch normalization follows each convolutional layer, with Rectified Linear Unit (ReLU) serving as the activation function. The architecture integrates two pooling layers and concludes with two fully-connected layers, containing 10 and 1 neurons respectively, producing the final output of the model. Figure ~\ref{fig:Fusion model}(a) shows the exact architecture of the adopted 3D-CNN. %\\[-0.3cm]

\paragraph*{SG-CNN}
\label{SG-CNN}
An SG-CNN effectively captures spatial information using a 2D graph representation, where each edge corresponds to a bond between atoms across all molecules. Both covalent and non-covalent bonds are encoded through two square $N \times N$ adjacency matrices $A_c, A_{nc}$ where  $A_{ij,c/nc}$ denotes the Euclidean distance (in angstroms \AA) between atom $i$ and atom $j$. Additionally, we establish two thresholds for covalent and noncovalent neighborhoods, $\alpha_c$ and $\alpha_{nc}$ respectively, such that $A_{ij,c} = 0$ if $A_{ij,c} \geq \alpha_c$ and $A_{ij,nc} = 0$ if $A_{ij,nc} \geq \alpha_{nc}$. In this study, the chosen thresholds were $\alpha_c = 1.5$\AA \ and $\alpha_{nc} = 4.5$\AA. For each molecule within the complex, spatial information and associated features are initially processed through a graph gated recurrent unit, incorporating information from its nearest neighbors. This stage, called the \emph{graph gather step}, is followed by four fully connected layers of size 12, 8, 6 and 1 respectively, ultimately yielding the SG-CNN final output. The exact architecture of the adopted SG-CNN is shown in Fig.~\ref{fig:Fusion model}(b).

\paragraph*{Quantum Fusion Model}
\label{quantum_fusion}
As depicted in Fig.\ref{fig:Fusion model}(c), the quantum fusion model combines outputs from the second-to-last layer of two CNN models, forming a 16-sized vector with 10 nodes from the 3D-CNN and 6 from the SG-CNN. This fusion optimally merges knowledge from the two CNNs, leading to superior performance. In contrast to the simple one-layer feed-forward neural network in Ref.\cite{ATOM}, our quantum fusion model integrates a QNN comprising a quantum circuit divided into two blocks. The first block, \emph{quantum encoding}, maps input data into a quantum circuit, while the second block, a \emph{parameterized quantum circuit (PQC)}, employs quantum operations to extract information from the encoded data.
%This will involve experiments with different quantum hardware providers, incorporating various error mitigation techniques, to comprehensively assess the method's effectiveness in NISQ devices.}

\paragraph*{Quantum Encoding Techniques}
A variety of quantum data encoding techniques have been recently developed in the literature \cite{hur2022quantum}. In this work, we focus on two of the most effective encoding techniques, analyzing and comparing their performance in the context of binding affinity prediction.

\begin{enumerate}
    \item \textbf{Amplitude encoding}, where the features are normalized and encoded in the amplitudes of the quantum state in the computational basis \cite{amplitudeEncoding}. This scheme requires only  $\lceil \log_2(n) \rceil $ qubits to encode a data sample into a quantum state, where $n$ represents the input dimension of the QNN. The depth of the embedding circuit grows as $O(poly(n))$ while the number of parameters subject to optimization scales as $O(log_2(n))$.
    \item \textbf{Hybrid Angle Encoding (HAE)}, where amplitude encoding is implemented using parallel blocks of independent qubits \cite{hur2022quantum}. The features are divided into $b$ blocks of size $2^m -1$, where $m$ is the number of qubits. Accordingly, $b \times m$ qubits are required to encode the whole data sample into a quantum state. 
\end{enumerate}

\paragraph*{PQC Architecture}
After encoding the data into qubits, they pass through a PQC, whose design is crucial to ensure efficient performance of the quantum fusion model \cite{yang2022bertquantum, qi2022classicalquantum}. A notable challenge is to choose an effective circuit that adequately represents the solution space while minimizing circuit depth and the number of parameters. Up to this point, two robust metrics have been introduced to assess the quality of PQCs \cite{sim2019expressibility}. The first metric, termed \textit{expressibility}, gauges the PQC's capacity to explore the Hilbert space, thereby generating a diverse array of quantum states \cite{sim2019expressibility}. %Expressibility is computed by analyzing the probability distribution of fidelity between two randomly chosen states from the circuit. This distribution is then juxtaposed with the random distribution of states (the Haar measure) \cite{sim2019expressibility}.
%, and the Kullback-Leibler divergence between the two distributions quantifies the circuit's expressibility \cite{sim2019expressibility}. A lower value of expressibility indicates a more expressive circuit. 
The second metric, termed \textit{entangling capacity}, quantifies the PQC's ability to generate entangled states \cite{meyer2002global}. %Among various entanglement measurement methods, we employ the Meyer-Wallach  measure \cite{meyer2002global} for its scalability and computational simplicity. 

In this work, we examine six distinct PQC architectures, shown in Fig. \ref{fig:circuits}, each composed of $L$ layers of quantum gates and characterized by differing levels of expressibility, entangling capacity, and number of training parameters. The goal is to identify the most optimized architecture for our quantum fusion model:
\begin{figure}[t!]
    \centering
\includegraphics[width=1.01\columnwidth]{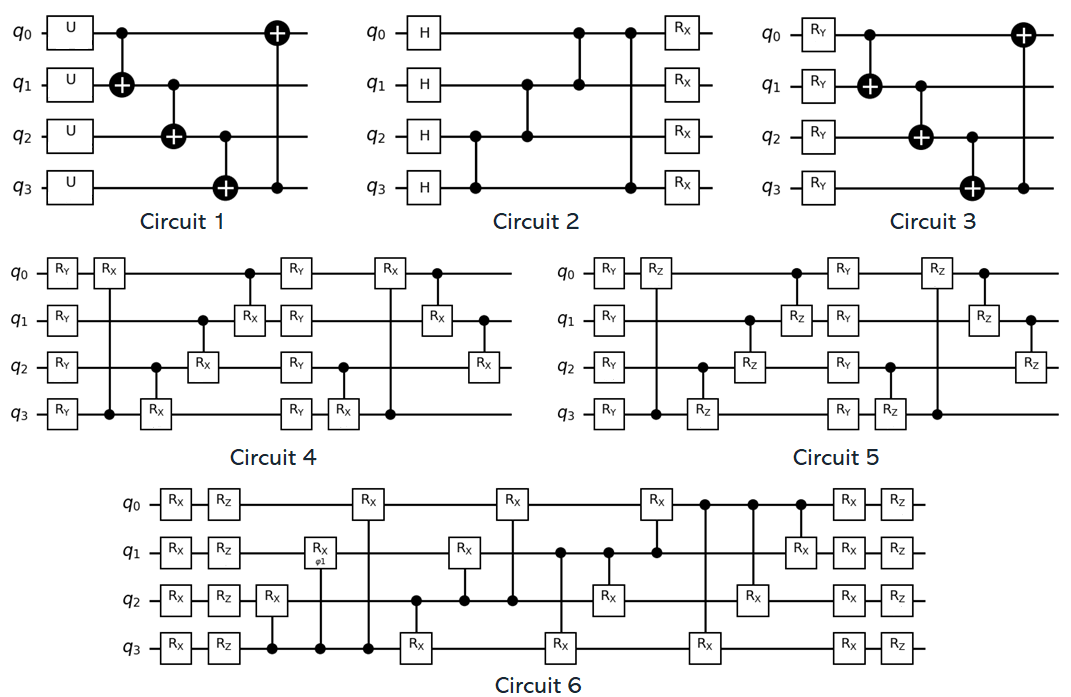}\vspace{-0.1cm}
    \caption{PQCs used as the ansatz of the quantum fusion model, labeled by their ID. Each figure represents a single layer of the final circuit. }\vspace{-0.5cm}
    \label{fig:circuits}
\end{figure}

\begin{itemize}
     \item \textbf{Circuit 1}. Comprised of 3D rotation gates with all-to-all CNOT connections forming \emph{strongly entangling layers}, this circuit demonstrates high entangling capacity and low expressibility. This circuit outperforms several other PQC architectures and classical counterparts in \cite{tuysuz2021hybrid}.
    \item \textbf{Circuit 2}.  Constructed with a layer of Hadamard gates, followed by all-to-all CZ connections and a layer of rotations, this PQC has high entangling capacity \cite{sim2019expressibility}.
    \item \textbf{Circuit 3}. Comprised of Y-axis rotations followed by a layer of CNOT gates, this PQC exhibits high entangling capacity and low expressibility, while incorporating fewer training parameters than Circuit 1.
    \item \textbf{Circuit 4}. Constructed with two layers of Y-axis rotations and two layers of controlled X-axis rotations, this PQC achieves a significant balance between entangling capacity and expressibility, but it entails higher complexity.
    \item \textbf{Circuit 5}. This PQC's structure is similar to Circuit 4, but with Z-axis rotations replacing the X-axis rotations.
    \item \textbf{Circuit 6}. Constructed with two layers of both X-axis and Z-axis rotations, along with two layers of all-to-all controlled rotations around the X-axis, this PQC has high expressivity, but also high complexity \cite{sim2019expressibility}.
\end{itemize}

%In this work, the architecture of the PQC is designed to generate a substantial level of entanglement and expressibility, by creating cross connections between the qubits. The PQC consists of several quantum gates which are controlled by classically-optimized parameters. Several PQC architectures were analyzed, and the optimal configuration identified consists of 3D rotation gates with all-to-all CNOT connections catalogued as \emph{strongly entangling layers}.

\vspace{-0.3cm} 
\subsection{Training Dataset}\vspace{-0.05cm}
\label{Data}
The PDBbind dataset~\cite{Pdbbind/site} (2020 version) is the input used to train the quantum fusion model. PDBbind represents an extensive compilation of experimentally determined binding affinity data between proteins and ligands. This dataset meticulously associates protein-ligand complexes with their respective affinity measurements, a curation process executed via manual extraction from peer-reviewed scientific publications. The latest PDBbind version, released in early 2020, contains a total of 19,443 protein-ligand complexes. Additionally, a subset of 5,316 samples has been compiled, specifically comprising high-quality complexes. Finally, an even higher-quality core set of 285 samples is derived. We utilize the refined set in the training and validation phases of our analysis, with 25\% of the data reserved for validation. Then, the core set is used for the testing phase. 

\vspace{-0.2cm}
\subsection{Data Pre-processing}%\vspace{-0.05cm}
\label{Preprocessing}
Before passing the PDBbind data into the two CNN architectures, we pre-process the raw data and extract pertinent input features using featurization techniques \cite{ATOM, domingo2023hybrid}. The input data to 3D-CNN consists of spatial representation of 3D structures of atoms of protein-ligand pairs. The atoms are voxelized into a grid of size $N\times N\times N$ with a voxel size of $1$ \si{\angstrom}, and the number of voxels set to $N=48$ to strike a balance between covering the entire pocket regions and lowering the input data size for the CNN model. $C=19$ features are extracted for each voxelized atom using OpenBabel \cite{vandermeersch2011open}, bringing the input data for 3D-CNN to a $C\times N\times N\times N$ matrix. Such features include the atom type, hybridization, number of heavy atom bonds, number of bonds with heteroatoms, structural properties, partial charge, and molecule type (protein vs ligand). Moreover, the input data to SG-CNN consists of a spatial graph representation of the protein-ligand complexes, with the atoms forming the nodes, and the bonds between atoms forming the graph edges \cite{ATOM}.

\vspace{-0.3cm}
\section{Quantum Error Mitigation Scheme}\vspace{-0.1cm}
\label{error_mitigation} \label{Error_Mitigation}
In this section, we introduce our tailored quantum error mitigation algorithm, which is evaluated with three quantum noise channels \cite{ZNE}. The first noise model, the \emph{amplitude damping channel}, characterized by a probability parameter $p$, simulates energy dissipation from a quantum state to its environment. The second model, the \emph{phase damping channel}, also with parameter $p$, mimics the loss of quantum information without energy loss. Finally, the \emph{depolarizing channel} introduces Pauli errors ($X$, $Y$, or $Z$) with equal probability $p$. 

\begin{figure}
     \centering
     \includegraphics[width=\columnwidth]{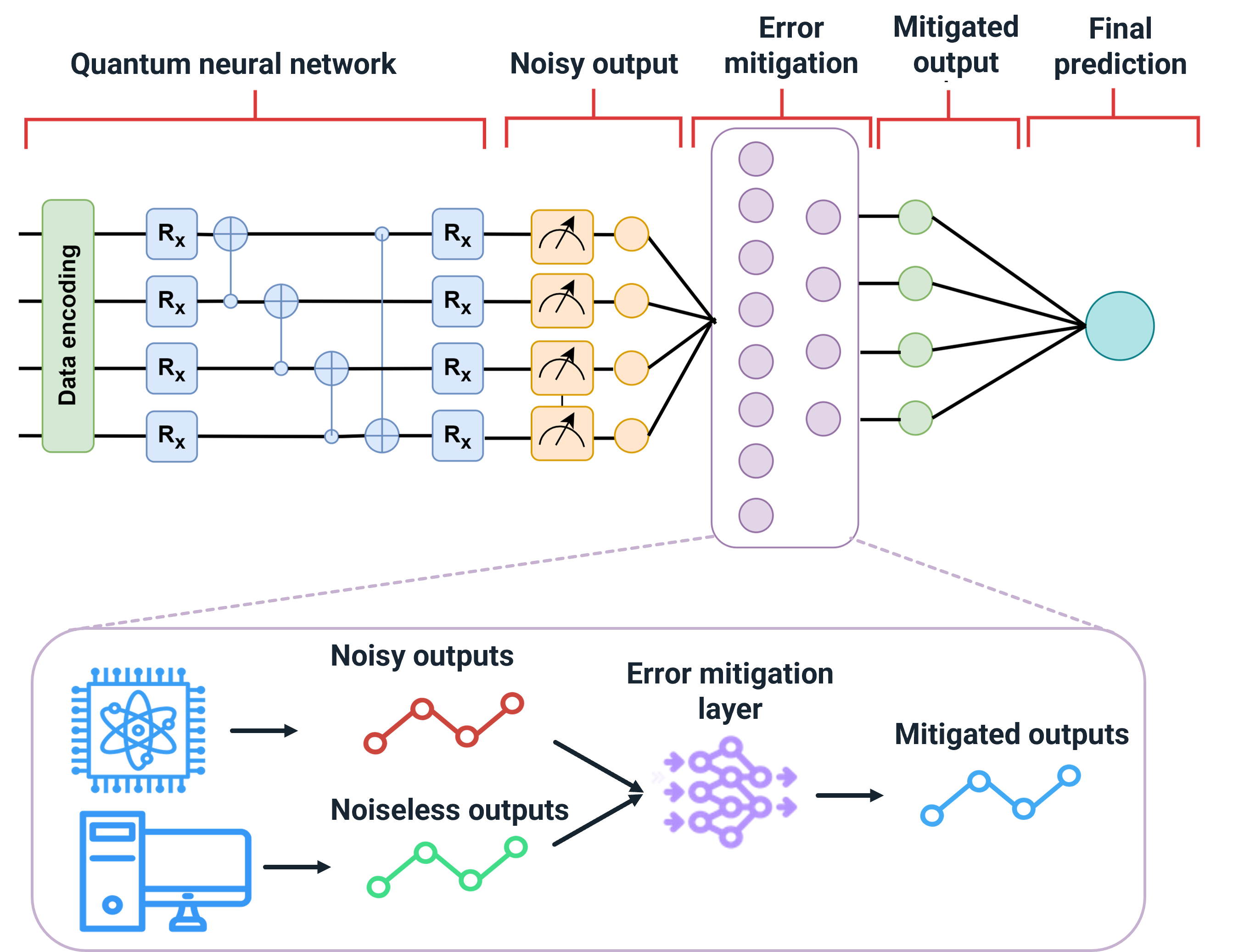}\vspace{-0.15cm}
     \caption{Proposed DREM algorithm architecture. The mitigated quantum fusion model uses an additional layer which performs error mitigation on the outputs of the noisy QNN.}
\label{fig:error_mitigation}\vspace{-0.35cm}
 \end{figure}

Our quantum error mitigation scheme involves training an ML model tailored to correct errors stemming from noisy quantum circuits. This proposed method, called \emph{data regression error mitigation} (DREM), is an extension of a prior work \cite{domingo2023hybrid} and has been customized specifically for our quantum fusion model. A schematic representation of our error mitigation algorithm is illustrated in Fig. \ref{fig:error_mitigation}. The mitigated quantum fusion model incorporates a DREM classical layer designed to mitigate errors stemming from the noisy QNN. This classical layer comprises two fully-connected layers, with dimensions 32 and 16, respectively. Prior to training the quantum fusion model, the DREM layer undergoes training to learn to perform error mitigation by processing examples of both noisy and noiseless outputs. The layer's parameters are fine-tuned by minimizing the mean squared error with $\mathcal{L}^2$ regularization between the noisy and noiseless outputs, with regularization parameter $\alpha$. 
%\begin{equation}
%\text{MSE}_R = 
%    \frac{1}{N_s} \sum_{i=0}^{N_s} \left[ \hat{y}_i - y_i \right]^2 
%    + \alpha ||W||^2,
%\end{equation}
%where $N_s$ is the number of samples in the training set, $\hat{y}_i$ is the $i$-th prediction of the DREM layer, $\alpha$ is the regularization parameter, and $||\cdot||$ is the $L^2$ norm.  
Once trained, the DREM layer is integrated into the mitigated quantum fusion model, as illustrated in Fig. \ref{fig:error_mitigation}. Subsequently, the DREM layer remains unaltered during the fusion model's training and operates solely as a corrective element for the noisy outputs. Consequently, it can be employed multiple times with no quantum overhead and minimal (classical) computational expense.

\vspace{-0.15cm}
\section{Simulations and Results} \vspace{-0.1cm}
\label{experiments}
Now, we conduct extensive simulations to compare the binding affinity prediction performance of our hybrid quantum fusion model against its classical counterpart using five metrics: root mean square error (RMSE), mean absolute error (MAE), $R^2$, Pearson coefficient, and Spearman coefficient. Optimal performance entails minimizing RMSE and MAE while maximizing 
$R^2$, Pearson, and Spearman coefficients.

%This section provides the details of the best QNN architecture for the quantum fusion model. 

%\vspace{-0.2cm}
\subsection{Experimental Setup}%\vspace{-0.1cm}
Following a comparative evaluation of various optimizers and hyperparameters, the training of both the classical and our proposed quantum fusion models is conducted using the ADAM optimizer with a learning rate of $\eta=0.002$, a batch size of 100 samples, and a mean squared error loss function. Moreover, in the QNN, $\sigma_{z}$ Pauli operator projection measurements are performed on all qubits to extract the classical measurement outcomes. The final prediction is made with a classical linear layer with a ReLu activation function. The optimal number of layers of the PQC is studied while considering varying depths of $L \in [1,14]$. In the proposed quantum error mitigation algorithm, the regularization parameter, $\alpha$, is set to $1 \times 10^{-5}$ in our simulations. The DREM layer is trained using 100 QNNs, each consisting of 10 layers from the circuit 1 architecture.

\vspace{-0.1cm}
\subsection{Binding affinity prediction performance}\vspace{-0.05cm}

% \begin{table}
% \begin{tabular}{| c | c | c | c | c | c|} 
%  \hline
%  Method & $r^2$ & MAE & Pearson & Spearman & RMSE \\ [0.5ex] 
%  \hline\hline
%  Class. Fsn & 0.60 & 1.05 & 0.777 & 0.766 & 1.37 \\ [0.5ex] 
%  \hline
%  Q'tum Fsn & 0.63 & 1.04 & 0.809 & 0.815 & 1.29 \\
%  \hline
% \end{tabular}
% \caption{Comparison of classical and quantum fusion model for the five error metrics.}
% \vspace{-0.5cm}
% \label{tab:performance}
% \end{table}

\begin{table}[t!] 
\caption{Performance of the classical and quantum fusion models.}\vspace{-0.2cm}
\label{tbl:Def}\footnotesize
\centering
\begin{tabular}{|p{1.3cm}|p{0.5cm}|p{0.6cm}|p{0.9cm}|p{1.1cm}|p{0.65cm}|}
 \hline
 Method & $R^2$ & MAE & Pearson & Spearman & MSE  \\ \hline
 Class. Fsn & 0.60 & 1.05 & 0.777 & 0.766 & 1.88 \\
   \hline
   Q'tum Fsn & 0.63 & 1.04 & 0.809 & 0.815 & 1.66 \\ 
   \hline
\end{tabular}\label{tab:performance}
\end{table}

%\vspace{-0.2cm}
\begin{table}[t!]\vspace{-0.2cm}
\caption{Quantum fusion model performance with different PQCs.}\vspace{-0.15cm}
\label{tab:performance_circuits}
    \centering\footnotesize
    \begin{tabular}{|p{0.2cm}|p{0.55cm}|p{0.55cm}|p{0.7cm}|p{0.94cm}|p{0.6cm}|p{0.95cm}|p{0.79cm}|}
        \hline
         ID & $R^2$ & MAE & Pearson & Spearman & MSE & Time (s) & \#params \\
        \hline
        1 & 0.63 & 1.04 & 0.809 & 0.815 & 1.78 & 116 & 125 \\
        2 & 0.546 & 1.085 & 0.763 & 0.781 & 2.14 & 88 & 45 \\
        3 & 0.508 & 1.104 & 0.728 & 0.738 & 2.32 & 66 & 45 \\
        4 & 0.600 & 1.047 & 0.791 & 0.792 & 1.883 & 215 & 165 \\
        5 & 0.591 & 1.055 & 0.78 & 0.779 & 1.93 & 228  & 165 \\
        6 & 0.607 & 1.04 & 0.802 & 0.799 & 1.85 & 436  & 285 \\
        \hline
    \end{tabular}\vspace{-0.4cm}
\end{table}

First, we compare the performance (in different metrics) of the classical and quantum fusion models, and summarize the results in Table~\ref{tab:performance}. We observe that the quantum fusion model provides a better performance in all five metrics, representing an up to 6\% improvement over the classical model.

\vspace{-0.1cm}
\subsection{Convergence of the quantum fusion model}\vspace{-0.05cm}
\begin{figure}[t]
    \centering
\includegraphics[width=0.9\columnwidth]{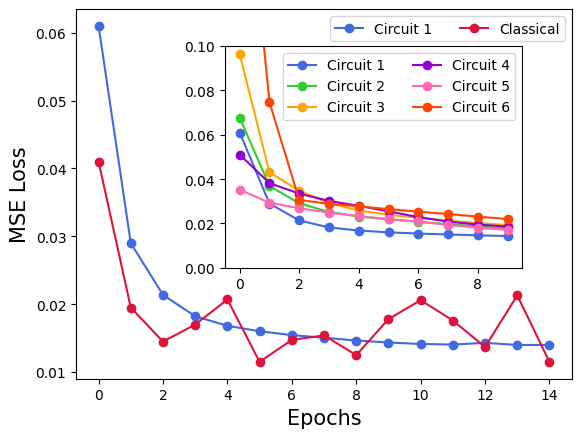}\vspace{-0.4cm}
    \caption{Loss function convergence for the classical and best quantum fusion models (with amplitude encoding, and 10 layers of PQC 1), together with the convergence of the six PQCs with 10 layers.}\vspace{-0.4cm}
    \label{fig:loss_profile}
\end{figure}
Next, in Fig.~\ref{fig:loss_profile}, we compare the convergence of the validation loss function for the quantum and classical fusion models. The convergence of the loss function for the quantum model is significantly smoother and more stable compared to the classical model. This can be attributed to the classical model having over three times the number of training parameters as the quantum model, which raises the likelihood of overfitting.  
\vspace{-0.3cm}
\subsection{Impact of the quantum architecture}
%The quantum fusion model's architecture plays a pivotal role in striking a balance between accuracy and time efficiency. In our initial experiment, we scrutinized the performance of the six quantum fusion models illustrated in Fig. \ref{fig:circuits}. 

\begin{figure}[t]
    \centering
    \includegraphics[width=0.85\columnwidth]{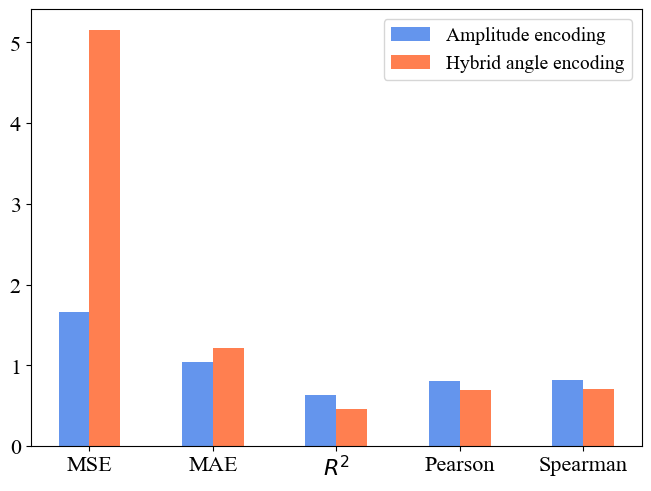}\vspace{-0.4cm}
    \caption{Test statistics resulting from two data encoding schemes and strongly entangling layers with $L=10$.}\vspace{-0.55cm}
    \label{fig:stats_QC}
\end{figure}
The inset in Fig. \ref{fig:loss_profile} depicts the convergence patterns of the six PQC architectures.  Table \ref{tab:performance_circuits} provides an overview of the performance of the six quantum fusion models, with $L=10$ layers, where Circuit 1 clearly achieves the best performance. While Circuits 2 and 3 train faster, they lack complexity necessary to for accurate exploration. This is in contrast with Circuits 4-6, which needed longer training times and achieved slightly worse performance due to their higher complexity that could lead to overfitting. Therefore, Circuit 1 is considered to achieve the best balance between complexity, expressibility, and entangling capacity.

Beyond the PQC design considerations, we compare amplitude encoding and HAE (with parameters $b=2$ and $m=4$) techniques using Circuit 1 with $L=10$. The results, depicted in Fig. \ref{fig:stats_QC}, clearly demonstrate the superior performance of amplitude encoding over the HAE technique. We also studied the impact of layer count on performance. Fewer layers result in higher errors due to limited Hilbert space access, while more layers increase overfitting risk. The optimal balance is found at $L=10$ layers.

\vspace{-0.2cm}
\subsection{Quantum error mitigation}
\begin{figure*}[!ht]
    \centering
\includegraphics[width=0.90\textwidth]{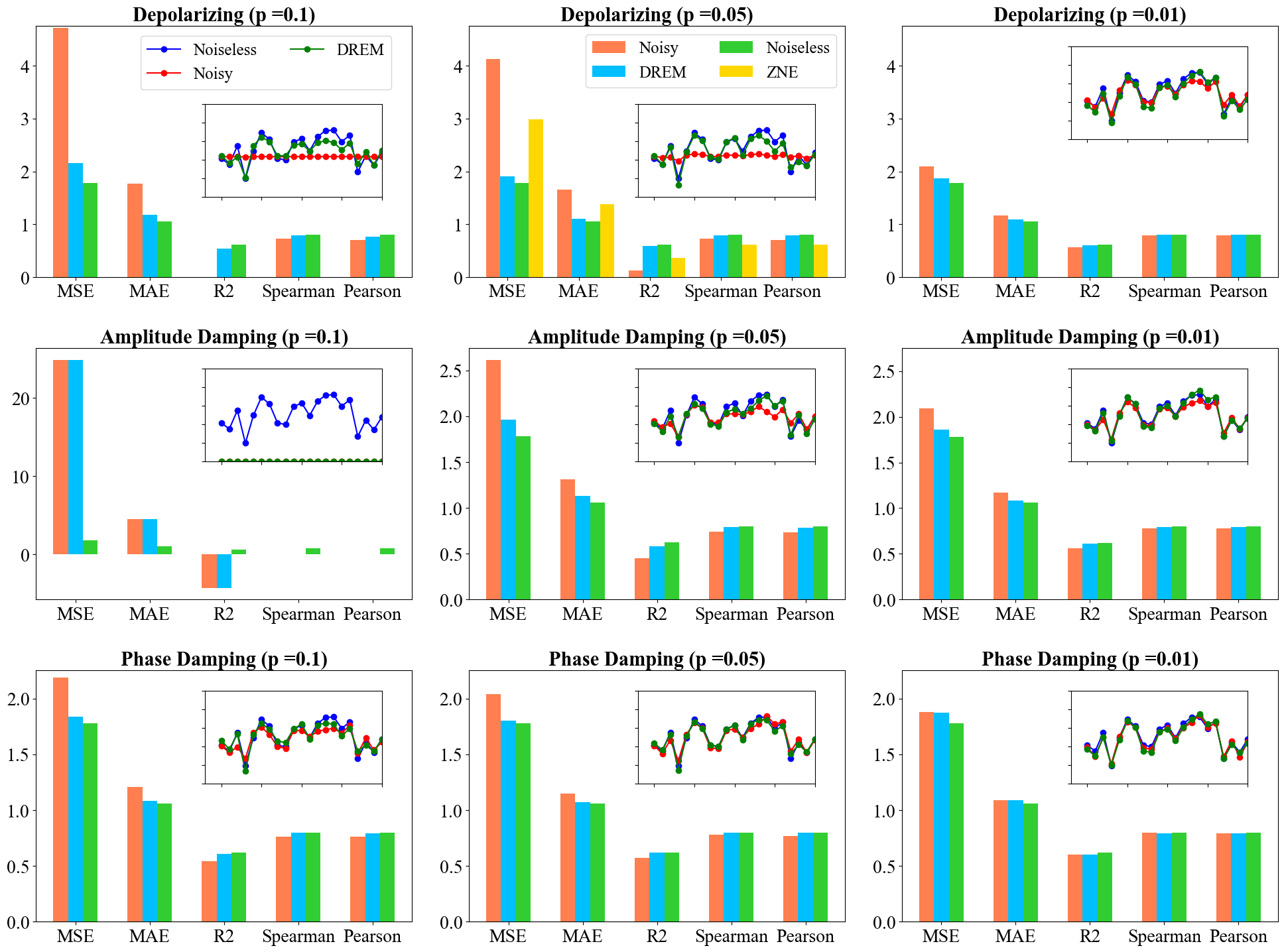}\vspace{-0.25cm}
    \caption{Performance of the DREM and ZNE algorithms evaluated using the five error metrics considered in this work, for different noise models and error rates. The insets contain the first 20 predictions of the quantum fusion models. Noiseless predictions are represented by blue curves, noisy predictions by red curves, and mitigated predictions by green curves. }
\label{fig:performance_noise}\vspace{-0.6cm}
\end{figure*}
%
%After analyzing the results of the noiseless quantum fusion models, we next study the performance of our proposed framework under practical noisy conditions and assess the obtained performance. In particular, we conduct simulations and compare the performance under three distinct quantum noise channels (see Sect. \ref{Error_Mitigation}) and assess their respective performances. 
Figure \ref{fig:performance_noise} illustrates the comparative performance of noisy, mitigated, and noiseless quantum fusion models, together with the sample predictions under various noise models and error probabilities. In particular, we consider three different quantum gate noise probabilities, $p=0.1, 0.05, 0.01$, which correspond to high, moderate, and low noise, respectively \cite{ZNE}, where the noise is applied after each quantum gate in the QNN.  All noise models reduce the variability of binding affinity predictions. Notably, at $p=0.1$, both depolarizing and amplitude damping noise channels cause the noisy model to predict a constant affinity. The phase damping noise channel exhibits slower variability reduction. The DREM algorithm effectively mitigates noisy QNN outputs for error rates $p\leq0.05$, restoring original outputs. The DREM method consistently improves error metrics across all error probabilities in the depolarizing channel, with the most significant enhancement at $p=0.05$. However, at $p=0.1$, there's a notable loss of information, and at $p=0.01$, the gap between noisy and noiseless metrics is less pronounced. For the amplitude damping channel, the DREM method effectively corrects errors up to $p=0.05$, but at $p=0.1$, information loss hampers its mitigation. Phase damping noise is effectively corrected for all probabilities by the DREM algorithm.

The DREM algorithm has been compared to the standard zero noise extrapolation (ZNE) \cite{ZNE} (with global folding and a scale factor of 3) for the depolarizing noise with $p=0.05$. As can be seen in Fig. \ref{fig:performance_noise}, the ZNE provides suboptimal results compared to the DREM. Moreover, the resulting ZNE model incurred a $3\times$ overhead in number of quantum gates, leading to a $10\times$ increase in training times. In contrast, our method introduced overhead only during the training of the DREM layer, representing a mere 10\% of the overall model training time. While the experiments presented in this work are based on quantum simulations, future works will include running the proposed model on actual quantum devices.

\vspace{-0.15cm}
\section{Conclusion}\vspace{-0.1cm}
\label{conclusion}
In this paper, we have proposed a novel hybrid QML model, offering accurate and reliable binding affinity predictions, necessary for drug discovery. The proposed model is a quantum fusion model that integrates 3D-CNN and SG-CNN classical models with a QNN. A thorough exploration of the circuit architecture and design has been performed, providing an optimal design of the quantum fusion model. Simulation results have validated the superiority of the quantum model relative to the classical fusion counterpart model, demonstrating a 6\% accuracy enhancement while achieving faster and smoother convergence, thus mitigating the risk of overfitting. Moreover, we have proposed an ML-based quantum error mitigation algorithm that outperforms existing error mitigation techniques. The proposed DREM algorithm provides strong error mitigation capabilities for error probabilities $p\leq 0.05$ without incurring additional quantum computational overhead.

%Moreover, our work also proves the potential effectiveness of the method with noisy real hardware aided with a relevant error mitigation technique. The results show that for error probabilities smaller than $p=0.1$, the DREM can succesfully recover the  prediction performance of the noiseless models.

\vspace{-0.15cm}
\section*{Code availability}\vspace{-0.1cm}
The code used in this work can be found at Ingenii's open access library: \href{https://github.com/ingenii-solutions/ingenii-quantum-hybrid-networks/}{https://github.com/ingenii-solutions/ingenii-quantum-hybrid-networks/}. %The data used for this study is publicly available at \href{http://www.pdbbind.org.cn/}{http://www.pdbbind.org.cn/} \cite{PDBBind}.

%\vspace{-0.2cm}
%\begin{spacing}{0.85}
\bibliographystyle{IEEEtran}
\bibliography{references}

\begin{thebibliography}{10}
\providecommand{\url}[1]{#1}
\csname url@samestyle\endcsname
\providecommand{\newblock}{\relax}
\providecommand{\bibinfo}[2]{#2}
\providecommand{\BIBentrySTDinterwordspacing}{\spaceskip=0pt\relax}
\providecommand{\BIBentryALTinterwordstretchfactor}{4}
\providecommand{\BIBentryALTinterwordspacing}{\spaceskip=\fontdimen2\font plus
\BIBentryALTinterwordstretchfactor\fontdimen3\font minus \fontdimen4\font\relax}
\providecommand{\BIBforeignlanguage}[2]{{%
\expandafter\ifx\csname l@#1\endcsname\relax
\typeout{** WARNING: IEEEtran.bst: No hyphenation pattern has been}%
\typeout{** loaded for the language `#1'. Using the pattern for}%
\typeout{** the default language instead.}%
\else
\language=\csname l@#1\endcsname
\fi
#2}}
\providecommand{\BIBdecl}{\relax}
\BIBdecl

\bibitem{quantummechanical}
V.~Govind~Kumar \emph{et~al.}, ``Binding affinity estimation from restrained umbrella sampling simulations,'' \emph{Nature Computational Science}, vol.~3, no.~1, pp. 59--70, 2023.

\bibitem{ddPNAS}
J.~Paggi \emph{et~al.}, ``Leveraging nonstructural data to predict structures and affinities of protein-ligan complexes,'' \emph{Proc. Nat. Acad. Sci.}, vol. 118, p. e2112621118, 2021.

\bibitem{vamathevan2019applications}
J.~Vamathevan \emph{et~al.}, ``Applications of machine learning in drug discovery and development,'' \emph{Nature reviews Drug discovery}, vol.~18, no.~6, pp. 463--477, 2019.

\bibitem{biamonte2017quantum}
J.~Biamonte, P.~Wittek, N.~Pancotti, P.~Rebentrost, N.~Wiebe, and S.~Lloyd, ``Quantum machine learning,'' \emph{Nature}, vol. 549, no. 7671, pp. 195--202, 2017.

\bibitem{debenedictis2018future}
E.~P. DeBenedictis, ``A future with quantum machine learning,'' \emph{Computer}, vol.~51, no.~2, pp. 68--71, 2018.

\bibitem{chehimi2024federated}
M.~Chehimi \emph{et~al.}, ``Federated quantum long short-term memory (fedqlstm),'' \emph{Quantum Machine Intelligence}, vol.~6, no.~2, pp. 1--13, 2024.

\bibitem{batra2021quantum}
K.~Batra \emph{et~al.}, ``Quantum machine learning algorithms for drug discovery applications,'' \emph{Journal of chemical information and modeling}, vol.~61, no.~6, pp. 2641--2647, 2021.

\bibitem{jimenez2018k}
J.~Jim{\'e}nez \emph{et~al.}, ``{K deep: protein--ligand absolute binding affinity prediction via 3d-convolutional neural networks},'' \emph{Journal of chemical information and modeling}, vol.~58, no.~2, pp. 287--296, 2018.

\bibitem{kwon2020ak}
Y.~Kwon \emph{et~al.}, ``{AK-score: accurate protein-ligand binding affinity prediction using an ensemble of 3D-convolutional neural networks},'' \emph{International journal of molecular sciences}, vol.~21, no.~22, p. 8424, 2020.

\bibitem{ATOM}
D.~Jones \emph{et~al.}, ``Improved protein–ligand binding affinity prediction with structure-based deep fusion inference,'' \emph{Journal of Chemical Information and Modeling}, vol.~61, no.~4, pp. 1583--1592, 2021.

\bibitem{mensa2023quantum}
S.~Mensa \emph{et~al.}, ``Quantum machine learning framework for virtual screening in drug discovery: a prospective quantum advantage,'' \emph{Machine Learning: Science and Technology}, vol.~4, no.~1, p. 015023, 2023.

\bibitem{domingo2023hybrid}
L.~Domingo, M.~Djukic, C.~Johnson, and F.~Borondo, ``Hybrid quantum-classical convolutional neural networks to improve molecular protein binding affinity predictions,'' \emph{arXiv preprint arXiv:2301.06331}, 2023.

\bibitem{he2016deep}
K.~He, X.~Zhang, S.~Ren, and J.~Sun, ``Deep residual learning for image recognition,'' in \emph{Proceedings of the IEEE conference on computer vision and pattern recognition}, 2016, pp. 770--778.

\bibitem{hur2022quantum}
T.~Hur \emph{et~al.}, ``Quantum convolutional neural network for classical data classification,'' \emph{Quantum Machine Intelligence}, vol.~4, no.~1, p.~3, 2022.

\bibitem{amplitudeEncoding}
I.~F. Araujo \emph{et~al.}, ``A divide-and-conquer algorithm for quantum state preparation,'' \emph{Scientific reports}, vol.~11, no.~1, p. 6329, March 2021.

\bibitem{yang2022bertquantum}
C.-H.~H. Yang \emph{et~al.}, ``When {BERT} meets quantum temporal convolution learning for text classification in heterogeneous computing,'' in \emph{IEEE International Conference on Acoustics, Speech and Signal Processing (ICASSP)}, 2022, pp. 8602--8606.

\bibitem{qi2022classicalquantum}
J.~Qi and J.~Tejedor, ``Classical-to-quantum transfer learning for spoken command recognition based on quantum neural networks,'' in \emph{IEEE Intl. Conf. on Acoustics, Speech, and Signal Processing}, 2022.

\bibitem{sim2019expressibility}
S.~Sim \emph{et~al.}, ``Expressibility and entangling capability of parameterized quantum circuits for hybrid quantum-classical algorithms,'' \emph{Advanced Quantum Technologies}, vol.~2, no.~12, p. 1900070, 2019.

\bibitem{meyer2002global}
D.~A. Meyer and N.~R. Wallach, ``Global entanglement in multiparticle systems,'' \emph{Journal of Mathematical Physics}, vol.~43, no.~9, pp. 4273--4278, 2002.

\bibitem{tuysuz2021hybrid}
C.~Tüysüz \emph{et~al.}, ``Hybrid quantum classical graph neural networks for particle track reconstruction,'' \emph{Quantum Machine Intelligence}, vol.~3, p.~29, 2021.

\bibitem{Pdbbind/site}
wwPDB consortium, ``{Protein Data Bank: the single global archive for 3D macromolecular structure data},'' \emph{Nucleic Acids Research}, vol.~47, no.~D1, pp. D520--D528, 2018.

\bibitem{vandermeersch2011open}
T.~Vandermeersch and G.~Hutchison, ``Open babel: An open chemical toolbox,'' \emph{Journal of Chemoinformatics}, vol.~3, no.~33, 2011.

\bibitem{ZNE}
Y.~Li and S.~C. Benjamin, ``Efficient variational quantum simulator incorporating active error minimization,'' \emph{Phys. Rev. X}, vol.~7, p. 021050, 2017.

\end{thebibliography}
%\end{spacing}

\end{document}